\documentclass[pss]{wiley2sp} 
\usepackage{graphicx}
\usepackage{amsmath}
\usepackage{ulem}
\usepackage{color}
\usepackage{cite}

\begin{document}
%

\titlerunning{Angular distributions of electrons from Na$_8$}

\title{Angular distributions of electrons emitted from free and
deposited Na$_8$ clusters }
\author{
Matthias\ B\"{a}r\textsuperscript{\textsf{\bfseries 1}},
Phuong Mai\ Dinh\textsuperscript{\textsf{\bfseries 2}},
Lyudmila\ V. Moskaleva\textsuperscript{\textsf{\bfseries 3}},
Paul-Gerhard\ Reinhard\textsuperscript{\Ast,\textsf{\bfseries 1,2}},
Notker\ R\"{o}sch\textsuperscript{\textsf{\bfseries 3}},
Eric\ Suraud\textsuperscript{\textsf{\bfseries 2}}
}

\authorrunning{Matthias\ B\"{a}r et al.}

\mail{e-mail
  \textsf{reinhard@theorie2.physik.uni-erlangen.de}, Phone:
  +49-9131-85-28462, Fax: +49-9131-85-28907}

\institute{%
  \textsuperscript{1}\,
  Institut f{\"{u}}r Theoretische Physik, Universit{\"{a}}t Erlangen,
Staudtstrasse 7, D-91058 Erlangen, Germany \\
  \textsuperscript{2}\,
Laboratoire de Physique Th\'{e}orique, IRSAMC, UPS and CNRS,
Universit\'{e} de Toulouse, 118 Rte de Narbonne, F-31062 Toulouse
cedex, France\\
  \textsuperscript{3}\,Department Chemie and Catalysis Research Center, 
Theoretische Chemie, Technische Universit\"{a}t
M\"{u}nchen, D-85748 Garching, Germany
}

\received{XXXX, revised XXXX, accepted XXXX} 
\published{XXXX} 


\abstract{
We explore from a theoretical perspective angular distributions of
electrons emitted from a Na$_8$ cluster after excitation by a short
laser pulse.  The tool of the study is time-dependent density-func\-tio\-nal
theory (TDDFT) at the level of the local-density approximation (LDA)
augmented by a self-interaction correction (SIC) to put emission
properties in order.  We consider free Na$_8$ and Na$_8$ deposited on
the surfaces 
MgO(001) or Ar(001). For the case of free Na$_8$, we
distinguish between a hypothetical situation of known cluster
orientation and a more realistic ensemble of orientations.  We also
consider the angular distributions for emission from separate
single-electron levels.  
%
} 
\maketitle

\section{Introduction}
\label{intro}

Optical methods have provided the key analyzing tools in cluster
physics over decades. In the first stage, optical absorption
measurements allowed one to collect rich information on structure and
dynamics of these small, nano-sized particles; for an overview see,
e.g., \cite{Bra93,Hee93,Kre93,Hab94a}. More information can be
gathered when one additionally measures the features of reaction
products in more detail. A most important channel in this context is electron
emission and thus there meanwhile exist many investigations that analyze
the properties of the electrons emitted after irradiation by a short
laser pulse. The first step in that direction is photoelectron
spectroscopy (PES) where the distribution of the kinetic energy of the
emitted electrons is recorded.  This method has been applied in
cluster physics since long, see e.g. \cite{McH89,Lic91}.  Stepping
further in refinement, one can determine the angular distribution of
the outgoing electrons which, in fact, is mostly done simultaneously
together with PES \cite{Pin99,Bag01,Ver04,Kos07a,Skr08,Bar09}.
Further challenging aspects come into play when considering clusters
in contact with substrates. 
That combination is often motivated (if not dictated) by
easier experimental handling. It is of great importance for possible
practical applications, and the effects at interfaces are also an
interesting problem for basic research, for up to date collections see
e.g. \cite{Mei00,Mei06}. It is obvious that the deposition on a
surface modifies the emission properties, in particular angular
distributions, thus calling for dedicated theoretical studies.
This paper is devoted to a theoretical exploration of angular
distributions of laser excited metal clusters, free and deposited on
insulating surfaces, MgO(001) and Ar(001).

{There is a broad choice of approaches for the description of
clusters on surfaces, from fully detailed quantum mechanical
treatments of all constituents \cite{Mos02a} to a robust jellium
treatment of cluster and interface \cite{Sem08a}. We are going here
for an intermediate strategy, detailed at the side of the highly
reactive cluster and less so for the inert substrate.}
As a basis of the description, we employ density-func\-tio\-nal theory
\cite{Dre90}, which provides a most efficient theoretical description
for the electronic structure and dynamics of clusters \cite{Rei03a}.
We simulate the detailed dynamics of laser excitation and subsequent
electron emission by time-dependent density-functional theory (TDDFT)
solved on a grid in coordinate space.  {The substrates are
inert and will be described classically in terms of
polarizable atoms \cite{Feh05a,Bae07a}.}  The computation of angular
distributions requires rather large numerical boxes. Therefore,
previous explorations dealt with approximations, a semi-classical
approach \cite{Gig03} which is confined to very intense laser pulse or
a quantum-mechanical, but two-dimensional axial, treatment
\cite{Poh03a,Poh04b} for lower intensities and short pulses.  Here we
are going for a fully three dimensional TDDFT analysis because both
simplifications are not applicable anymore for the studies intended
here and because the progress of computational resources meanwhile
allows a TDDFT analysis in full three spatial dimensions.  We will
take Na$_8$ as a test case and compare free Na$_8$ with Na$_8$ on
MgO(001) as well as Na$_8$ on Ar(001).

The study concentrates on direct electron emission, i.e. those
electrons which are leaving the cluster with or directly after laser
impact. In practice, there is a competition between direct and thermal
emission. The preferred exit channel depends on the length of the
laser pulse. For pulses longer than the collisional relaxation time
induced by two-electron processes, thermal emission dominates,
while shorter pulses change the weight towards direct emission
\cite{Cam00,Sch01,Poh04a,RMP08}. We will use here pulses with full
width at half maximum of 20 fs, which surely are at the side of
{dominant} direct emission.

There is also a subtle point about the orientation of the cluster
relative to the laser polarization. Experiments with clusters in the gas
phase deal, in fact, with an equi-distributed ensemble of orientations,
while deposited clusters have a well defined orientation due to the
known surface structure. We will briefly address this question,
considering both free clusters with known orientation and averaging
over cluster orientations.
We will then return to oriented clusters and
investigate the changes caused by deposition on a surface.
We will discuss how the specific angular distributions for
emission of each single-electron state separately combine to the total
distributions and how this helps to understand the pattern observed in emission
from deposited clusters.  Furthermore, we will investigate and compare
two different substrates, MgO(001) and Ar(001) surfaces. Both are
insulators with a rather large band gap. The  MgO(001) surface shows
more corrugation and has a stronger interface attraction, while
Ar(001) is softer and dominated by core repulsion.

The paper is outlined as follows.
In section \ref{sec:model}, we briefly summarize the formal framework,
numerical strategies, and the basic properties of the test cases.
In section \ref{sec:free}, we present and discuss results on free
Na$_8$ and also explore the double differential distributions.
Section~\ref{sec:depos} compares results for free Na$_8$ to those for
Na$_8$ deposited on MgO(001) and Ar(001).
Conclusions are given in section \ref{sec:concl}.

\section{Brief summary of the model}
\label{sec:model}

\subsection{The degrees of freedom}

The hierarchical quantum-mechanical-molecular-mecha\-ni\-cal
(QM/MM) model has been described in detail elsewhere 
(see \cite{Bae07a,Bae08a} and \cite{Din09} for a recent detailed
review).  Here we give a brief outline.  The constituents of the
system and their degrees of freedom are:
$\varphi_{n}(\vec{r},t)\,,\,n=1,...,N_{\mathrm{el}}$ for valence electrons of the
Na cluster,
$\vec{R}_{i^{\mathrm{(Na)}}}\,,\,i^{\mathrm{(Na)}}=1,...,N_{\mathrm{i}}$ 
for the positions of the Na$^{+}$ ions,
$\vec{R}_{i^{(c)}}\,,\,i^{(c)}=1,...,M$  for the positions of the O
cores,
$\vec{R}_{i^{(v)}}\,,\,i^{(v)}=1,...,M$ for the centers of the O valence
clouds,
$\vec{R}_{i^{(k)}}\,,\,i^{(k)}=1,...,M$ for the positions of the
Mg$^{2+}$ ions.
The Na cluster is treated in standard fashion: The valence electrons
quantum-me\-cha\-ni\-cal\-ly and the ions classically \cite{Rei03a,Cal00}.
The MgO substrate is composed of two species~: Mg$^{2+}$ cations and
O$^{2-}$ anions. The cations are electrically inert and can be treated
as charged point particles; they are labeled by $i^{(k)}$. The anions
are easily polarizable; this fact is accounted for by associating {them}
with two constituents~: A valence electron distribution (labeled by
$i^{(v)}$) and the complementing core (labeled by $i^{(c)}$). All ions
of the MgO substrate are described as classical degrees of freedom in
terms of positions $\vec{R}_{i^{\mathrm{(type)}}}$.  The difference
$\mathbf{R}^{(c)} -\mathbf{R}^{(v)}$ describes the electrical dipole
moment of an O$^{2-}$ anion.
The Mg and O ions are arranged in crystalline order corresponding to
bulk MgO. The dynamical degrees of freedom for Mg and O are taken into
account in an active cell of the MgO(001) surface region underneath
the Na cluster. The active cell is embedded in an ``outer region'' of MgO
material where only ions are kept fixed, while oxygen dipoles remain
fully dynamical. Beyond that region,
only the Madelung potential is considered.  The effect of the outer
region on the active part is given by a time-independent shell-model
potential taken over from \cite{Nas01a}.
Actually, the substrate consists of six layers each containing 784
Mg$^{2+}$ ions and 784 O$^{2-}$ ions.  The ions in the lowest layer are
fixed to prevent {them} from relaxing and forming an artificial second
surface. 
The active cell consists of three layers, each
containing square 
arrangements of 242 Mg$^{2+}$ cations and 242 O$^{2-}$ anions.

In the case of Ar, the modeling follows a similar, although
simplified, track~\cite{Feh05a,Feh05c}. The Ar substrate is composed of
only once species, neutral Ar atoms, each of which being characterized
by its position and dipole moment (exactly as O$^{2-}$ anions). The substrate
comprises 384 atoms; this model has been shown to provide a fair
approximation to bulk \cite{Din07b,Din08a}.

\subsection{Energy and fields}

The total energy is decomposed as
$
E=E_{\mathrm{Na}}+E_{\mathrm{Surf}}+E_{\mathrm{coupl}}
$
where $E_{\mathrm{Na}}$ describes an isolated Na cluster,
$E_{\mathrm{Surf}}$ the MgO(001) or Ar (001) substrate, and
$E_{\mathrm{coupl}}$ the coupling between the two subsystems. For
$E_{\mathrm{Na}}$, we take the standard {functional of TDDFT at
the level of the local density approximation (LDA)}, {coupled
with molecular dynamics (MD),} as in
previous studies of free clusters \cite{Rei03a,Cal00} including an
average {density} self-interaction correction \cite{Leg02}. The
energy of the 
substrate and the coupling to the Na cluster consists of the
long-range Coulomb energy and some short-range repulsion which is
modeled through effective local core-potentials
\cite{Nas01a,Rez95,Dup96}.  To avoid the Coulomb singularity and to
simulate the finite extension of Ar, Mg$^{2+}$ and O$^{2-}$ ions, we
associate a smooth charge distribution
$\rho(\vec{r})\propto\exp{(-{\vec{r}^{2}}/{\sigma^{2}})}$ with each of
these ionic centers. This yields a soft Coulomb potential to be used
for all active particles.  The model parameters were calibrated to
represent results of calculations where a fully quantum-mecha\-ni\-cal
description of the active MgO cell was employed in the case of MgO,
for details see \cite{Bae07a}. The model parameters for the Ar
substrate were chosen to reproduce model calculations of the NaAr
dimer~\cite{Gro98}, in turn fitted to ab-initio data.

The field equations obtained by variation of the above energy are
augmented by the external laser field
\begin{equation}
  U_\mathrm{las}
  =
  -e \, \mathbf{r}\!\cdot\!\mathbf{n}_\mathrm{pol} \, E_0
  \sin^2\left(\frac{t}{T_\mathrm{pulse}}\pi\right)
  \sin\left(\omega_\mathrm{las}t\right)
\end{equation}
which is activated only in the time interval $0\leq t\leq
T_\mathrm{pulse}$.  The field strength $E_0$ is related to the
intensity $I$ as 
$I/(\mathrm{W}\,\mathrm{cm}^{-2})
=
27.8\left[E_0/(\mathrm{V}\,\mathrm{cm}^{-1})\right]^2
$.
The total pulse length $T_\mathrm{pulse}$ corresponds to a
full width at half (intensity) maximum as FWHM$\approx
T_\mathrm{pulse}/3$.  The laser polarization
$\mathbf{n}_\mathrm{pol}$ can take any direction. We will consider
$\mathbf{n}_\mathrm{pol}=\mathbf{n}_z$, the direction perpendicular to
the surface, i.e. along the symmetry axis of Na$_8$, and one direction
orthogonal to it with $\mathbf{n}_\mathrm{pol}=\mathbf{n}_x$ (for the
geometry, see section \ref{sec:struct}). A laser with polarization
$\mathbf{n}_z$ is, of course, an idealization because it would
correspond to a plane wave running parallel to the surface, but it
serves to model rather flat impact as compared to the
perpendicular impact of $x$-polarization.

From the energy functional, once established, one derives the static
and dynamical equations variationally in a standard manner
\cite{Rei03a}.

\subsection{Solution scheme}

The numerical solution of the coupled quantum-classical system
proceeds as described in \cite{Bae07a,Feh05a,Feh05b}. The electronic
wave functions
and the spatial fields are represented on a Cartesian grid in
three-dimensional coordinate space. 
The numerical box employed here has a size of $(64\, a_0)^3$.
The spatial derivatives are
evaluated via fast Fourier transformation. The ground state
configurations were found by accelerated gradient iterations for the
electronic wave functions \cite{Blu92} and simulated annealing for the
ions in the cluster and the substrate. Propagation is done by the
time-splitting method for the electronic wave functions \cite{Fei82} and
by the velocity Verlet algorithm for the classical coordinates of
Na$^{+}$ ions and MgO {or Ar} constituents.

\subsection{Gathering angular distributions}

Electrons emitted from the cluster will eventually arrive at the
boundaries of the box. In order to suppress re-feed of these electrons
back into the box, we employ absorbing boundary conditions
\cite{Cal00,Ull00b}.
This is achieved by the mask function ${\cal M}(\vec r)$ defined as~:
\begin{subequations}
\begin{equation}
  {\cal M}(\vec r) 
  =
  \left\{\begin{array}{lcl}
    1 & \mbox{for} & |\vec r|<R_\mathrm{cut}
  \\[2pt]
    \left[
    \sin\left(\displaystyle
     \frac{R_\mathrm{box}\!-\!|\vec r|}
          {R_\mathrm{box}\!-\!R_\mathrm{cut}}
     {\frac{\pi}{2}}    
    \right)\right]^{1/4}
    & \mbox{for} & R_\mathrm{cut}\leq |\vec r| \leq  R_\mathrm{box}
  \\[2pt]
    0 & \mbox{for} & |\vec r|>R_\mathrm{box}
   \end{array}
   \right.
\end{equation}
where $R_\mathrm{cut}$ is the cut-off radius outside which absorption
starts and $R_\mathrm{box}$ is the minimal radial distance from the
origin to the closest point on the boundaries.
The Kohn-Sham time step actually performed with the time-splitting
method \cite{Fei82} is thus augmented by an absorbing step as
\begin{equation}
\tilde\varphi=\hat U^{TV}\varphi(t) \quad
\rightarrow
\quad
\varphi(t+\delta t)={\cal M}(\vec r)\tilde\varphi
\end{equation}
\end{subequations}
where $\hat U^{TV}$ is the unitary propagation operator.  Applying the
mask function ${\cal M}$ to the orbitals gently removes density
approaching the box boundary and prevents it from being reflected.

To compute the angular distribution of emitted electrons,
the absorbed density is accumulated for each state and
each (absorbing) grid point as
\begin{subequations}
\begin{equation}
{
  \Gamma_i(\vec r)
  =
  \int_0^\infty \textrm dt\,
  \left|
  \left[1-{\cal M}(\vec r)\right]\,\hat U^{TV}\varphi_i(t)
  \right|^2
  \;.
}
\end{equation}
By definition of $\cal M$, the field $\Gamma(\vec r)$ is non-vanishing
only in {the spherical absorbing zone}. The angular
distribution of emitted electrons is finally gathered by dividing the
{absorbing zone} into radial segments $A_\nu$, and integrating
$\Gamma_i(\vec r)$ over those segments.
The Photoelectron Angular Distribution (PAD) then becomes
{
\begin{eqnarray}
  \qquad
  \frac{\textrm dN_{\rm esc}^{i}(\theta,\phi)}{\textrm d\Omega}
  &\propto& \frac{1}{||A_\nu(\theta,\phi)||}
  \int_{A_\nu} \textrm d^3 \vec r\, \Gamma_i(\vec r)
  \;,
\label{eq:crossecti}
\\
  \frac{\textrm dN_{\rm esc}(\theta,\phi)}{\textrm d\Omega}
  &=&
  \sum_{i=1}^{N_\mathrm{el}}
  \frac{\textrm dN_{\rm esc}^{i}(\theta,\phi)}{\textrm d\Omega}
\label{eq:crossect}
\end{eqnarray}
}
\end{subequations}
where $||A_\nu(\theta,\phi)||$ denotes the area of the segment $A_\nu$
on the surface of a unit sphere. 
{Eq. (\ref{eq:crossect}) provides
the total cross-section, while Eq. (\ref{eq:crossecti}) the cross-section
for emission from a specific quantum state $\varphi_i$.}

We will sometimes present simpler polar distributions, averaging the
full distributions $\textrm dN_{\rm esc}(\theta,\phi)/\textrm d\Omega$
over $\phi$ and dividing by the polar volume element $\sin\theta$
{(see Figure \ref{fig:struct}).} This is convenient when the variation
in $\phi$ is weak. However, even if the azimuthal distribution may
carry interesting details, particularly for deposited clusters, we
will see that the reduced view can bring some useful information.

{It is to be noted that we consider here PAD which are
  integrated over all outgoing electron momenta. 
  The state-specific PAD carry automatically some information
  on the electron spectra because the dominant emission strength
  goes into the first (multi-)photon peak above continuum threshold.
  More detailed information can be gathered
  with techniques as
  proposed in \cite{Poh03a,Poh00}. They also allow to produce a double
  differential cross-section providing the photo-electron spectra in
  each angular bin separately. An explanation of this technique and
  the much more complex analysis of the emerging data is postponed
  to a later publication. 
}

\subsection{Structure of the test cases}
\label{sec:struct}

As test cases, we will consider the free Na$_8$ cluster and Na$_8$
deposited on MgO(001) or Ar(001) surfaces.  Starting point of the
laser-induced dynamics is a well relaxed structure of each
system. These structures had been discussed extensively in
\cite{Bae07a}.
\begin{figure}[htbp]
\centerline{
\includegraphics[width=0.65\linewidth]{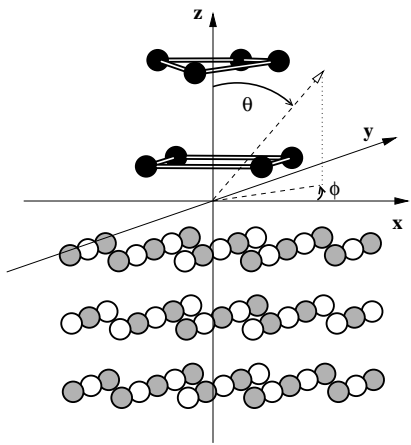}
}
\caption{\label{fig:struct}
The structure of Na$_8$ on MgO(001). Na ions are indicated by black
spheres, O ions by white ones, and Mg ions by gray ones. 
The structure of Na$_8$ on Ar(001) is essentially similar.
The bond distance in the two four-fold rings of Na$_8$
is 6.2 $a_0$ and the distance between the  two rings is 5.8 $a_0$.
 The equilibrium
distance between the lower cluster plane and
the first surface layer is 5 $a_0$.}
\end{figure}
Figure \ref{fig:struct} illustrates the structure of Na$_8$ on
MgO(001).  The symmetry axis of Na$_8$ which coincides for the
deposited Na$_8$ with the axis perpendicular to the surface is taken
as $z$-axis.
The MgO surface is a cut from the cubic crystal structure. On the
surface, the O and Mg ions are arranged in squares where next
neighbors are always the other species. 

The Na$_8$ cluster has a
highly symmetric configuration out of two rings of four ions each,
twisted relative to each other by 45$^\circ$ to minimize the Coulomb
energy. The Na$_8$ is rather rigid and only very little affected by
the surface. Free Na$_8$ is very similar to the deposited
cluster shown here, with bond lengths differing by less than
3\%. The same holds for Na$_8$ on Ar(001).
\begin{figure}[htbp]
\centerline{
\includegraphics[width=80mm,angle=0]{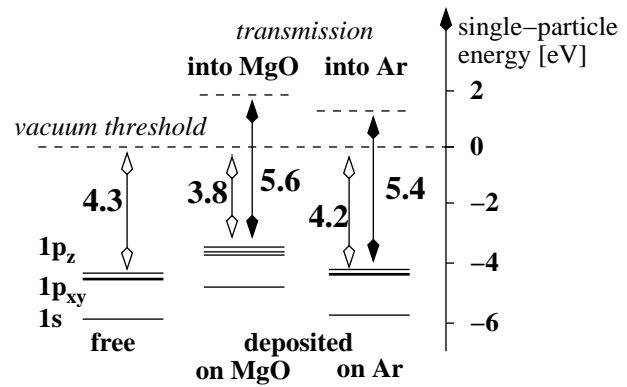}
}
\caption{\label{fig:Na8-spe}
The single-electron levels of free Na$_8$ and
of Na$_8$ deposited on MgO(001) or Ar(001).
The degeneracy of the $1p_x$ and $1p_y$ levels in free
Na$_8$ is indicated by a heavier line.
The vacuum threshold is at zero energy.
The IPs are indicated by vertical lines with open arrows
and the transmission barriers by vertical lines with
filles arrows. 
The numbers beneath are the IPs or barriers in units of eV. 
}
\end{figure}
The electronic cloud of free Na$_8$ exhibits close
to spherical shape because $N_\mathrm{el}=8$ electrons corresponds to
a strong  electronic shell closure for Na clusters~\cite{Bjo99}. That changes for
the deposited cluster.  The surface destroys the reflection symmetry
which, in turn, mixes single-electron states of different
$z$-parities. The presence of the surface does also shift the
single-electron levels and the ionization potentials (IP).
%
The relations are sketched in figure \ref{fig:Na8-spe}.  The IP is
$4.3$~eV for  free Na$_8$, $3.8$~eV for Na$_8$ on MgO(001) and $4.2$
~eV on Ar(001).  The transmission barrier for emission from deposited
Na$_8$ into the substrate lies at $+5.6$~eV for MgO(001) and $+5.4$~eV for
Ar(001): both are substantially  larger than the IP. The $1p_x$ and
$1p_y$ levels of free Na$_8$ are perfectly degenerate, slightly
split from $1p_z$ due to a very small quadrupole deformation of
Na$_8$. The ionic structure of  Na$_8$ deposited on MgO(001) hardly
changes because the magic electron configuration renders that cluster
very rigid.
There is an overall up-shift due to core repulsion from MgO.  The
symmetry breaking due to the surface orientation slightly splits the
$1p_{x,y}$ levels, by 0.204~eV.  Much less shifts are seen for Na$_8$
on Ar(001) and the splitting between $1p_x$ and $1p_y$, 0.05~eV, is
also much smaller. That difference correlates to the much lower
interface energy of the Na-Ar system. Indeed, the bottom layer of
Na$_8$ is within 5.1 $a_0$ from the MgO(001) surface, much closer to
the substrate than in the case of Ar(001) (6.4 $a_0$).

\begin{figure}[htbp]
\centerline{
\includegraphics[width=70mm,angle=0]{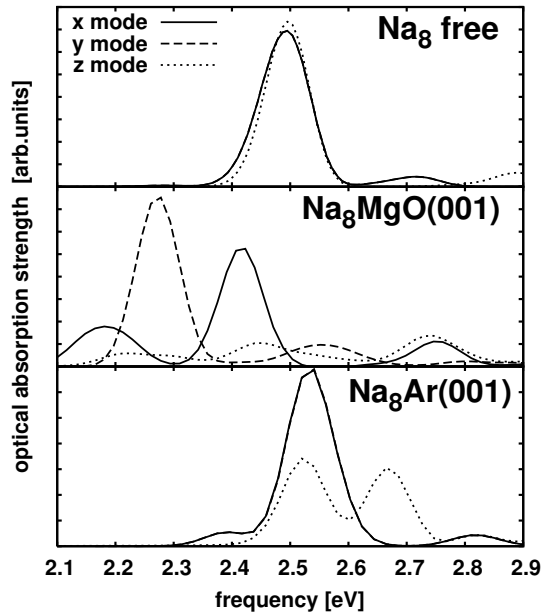}
}
\caption{\label{fig:Na8-substr-opt}
{
The optical absorption spectrum (dipole response)  
of free Na$_8$ and
of Na$_8$ deposited on MgO(001) or Ar(001).
The spectra from the modes in the three basic directions
are shown separately as indicated. The $z$ axis stands for
the direction perpendicular to the surface and along
the symmetry axis of Na$_8$.}
}
\end{figure}
{ The optical absorption spectrum of our test cases is shown in
figure \ref{fig:Na8-substr-opt}.
One recognizes} the pronounced Mie surface plasmon
resonance \cite{Bra93,Kre93,Rei03a}. For free Na$_8$, there is one
clean peak at 2.45~eV almost the same for each mode. The surfaces lead
to some spectral fragmentation, about 1.4~eV for MgO(001) and 0.2~eV
for Ar(001).  The highly polarizable MgO(001) also induces a small
down-shift of the resonance center to 2.31~eV while core repulsion
outweighs polarization for Ar(001) leading to a small up-shift to
2.54~eV. {The MgO(001) surface leads to a strong fragmentation
of the $z$ mode essentially due to symmetry breaking \cite{Bae07a}.}

In the following studies, we will vary laser frequency
$\omega_\mathrm{las}$ and polarization. We work in all cases with the
same pulse length $T_\mathrm{pulse}=60$ fs, which corresponds to
FWHM$=20$ fs. 
The angular distributions in the high-intensity regime are always
concentrated on forward and backward emission along the laser
polarization axis, whereas a high sensitivity to all details of the
excitation is found in the regime of weak perturbations \cite{Poh04b}.
The present study thus aims at staying close to the perturbative regime
to explore the rich variety of distributions. The intensity should be
low enough for a perturbative regime being valid
\cite{Poh03a,Poh04a,Fai87}, but also sufficiently high to provide signals
safely above numerical noise. We found an intensity of $I=10^9$
W/cm$^2$ to be a good compromise and we used that value in most of the test
cases. Use of a different value  will be clearly indicated.

\section{Brief survey of free Na$_8$}
\label{sec:free}


Detailed studies of the PAD of free clusters {will be discussed
in a forthcoming publication}. We briefly summarize here the results.

Variation of the laser intensity shows almost constant pattern of the
PAD throughout the regime of moderate intensities. For further
increasing intensities, the structures change steadily towards simple
forward-backward scattering.
This is related to the transition from the
frequency-dominated regime of moderate intensities to the
field dominated highly non-linear regime
\cite{RMP08,Cal00,Rei99a,Zwi99}.

Variation of frequency leads to dramatic changes in the angular
distributions.  
\begin{figure}[htbp]
\centerline{
\includegraphics[width=77mm]{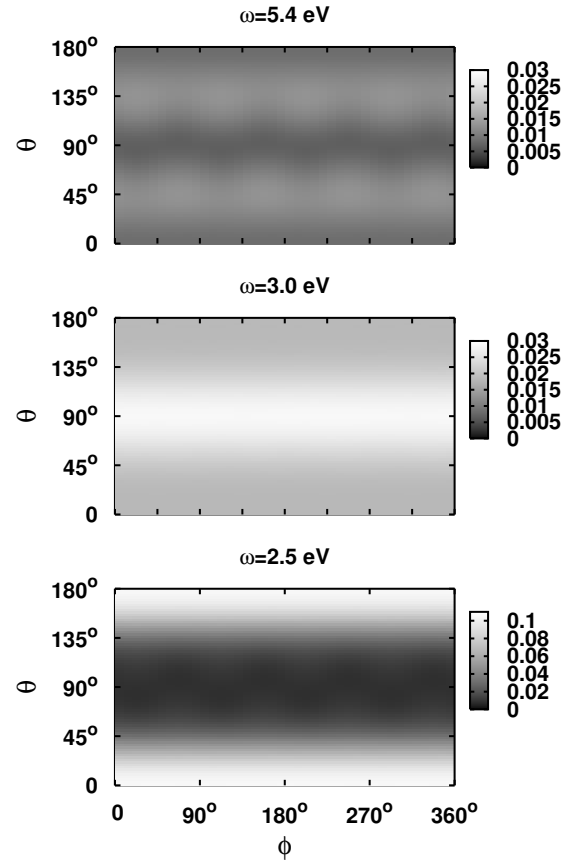}}
\caption{\label{fig:angdist-frequency}
Gray scale plot of angular distributions for free Na$_8$ excited with
three different laser frequencies, as indicated.
The gray scale is used in arbitrary units.
High emission is signified by white and no emission correlates
to deep black.
The frequencies had been selected to display the three
different patterns which could be found. 
The laser is polarized along the $z$-axis, the symmetry axis of Na$_8$.
}
\end{figure}
They vary amongst three typical patterns. 
{
We have chosen three frequencies to have an example for each
type. Results}
are shown in figure \ref{fig:angdist-frequency}.
The most common case is forward scattering (along the axis of
polarization, $\theta=0,\pi$), seen here in the lowest panel for
$\omega_\mathrm{las}=2.5$ eV.
Sometimes one encounters ``diagonal scattering''
where electron emission
is concentrated on a double cone around angle $\pi/4$ (with
respect of the $z$-axis), see
upper panel for $\omega_\mathrm{las}=5.4$ eV. 
Sideward scattering as exemplified in the middle panel, is observed
occasionally, here for $\omega_\mathrm{las}=3.0$ eV.
{Note that the spectral relations of these three
cases are very different.
For 5.4 eV, we have a one-photon process for the $1p$ states
but a (much suppressed) two-photon process from the $1s$ state.
For 3 eV, we have a two-photon process for both, the $1p$ and
the $1s$ shell. And for 2.5 eV, we have a two-photon emission
from the $1p$ while the $1s$ shell, again, requires one more
photon.  
}

The strong frequency dependence of the PADs can be related to  the
fact that the
strongest emitting single-electron state
depends sensitively on the relationship between laser frequency and IP,
and that the emission from each state looks very differently.
This is demonstrated by analyzing
the PADs of individual single-electron states.
Experimentally, this is achieved  by measuring the PAD simultaneously with
the photoelec\-tron spectrum (PES)~\cite{Bag01,Kos07a,Kos07b}.
The energy selection by PES allows one to associate the PAD to
the single-electron states from which these were produced.
Theoretical calculations have the separate information trivially
at hand, as seen in Eq.~(\ref{eq:crossect}). 
\begin{figure}[htbp]
\centerline{
\includegraphics[width=70mm]{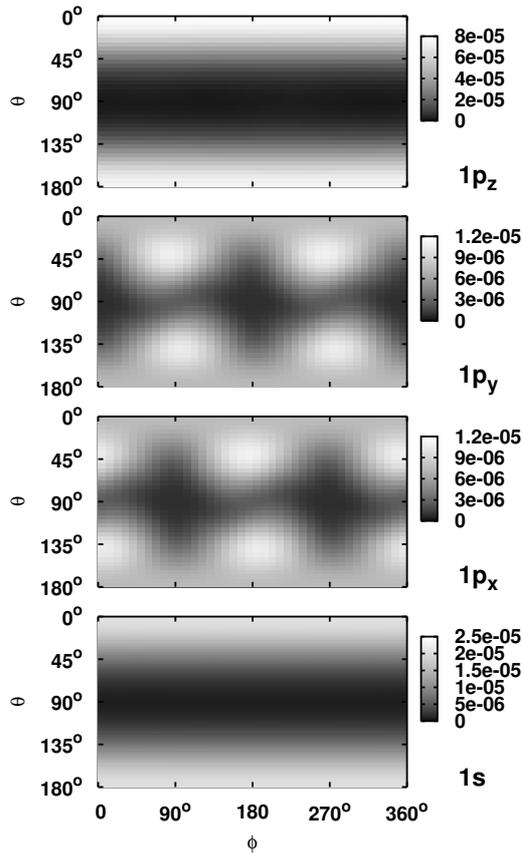}}
\caption{\label{fig:ang8orbital_019}
State selective angular distribution
 $dN_{{\rm esc},i}(\theta,\phi)/d\Omega$
of electrons which were emitted from
single-electron state $\varphi_i$ of free Na$_8$ as indicated,
given in arbitrary units.
The laser was polarized along the $z$ axis and
had a frequency of $\omega_\mathrm{las}=2.5$ eV.
}
\end{figure}
Figure \ref{fig:ang8orbital_019} shows the state-specific PAD of free
Na$_8$ for the (hypothetical) case that the clusters symmetry axis is
aligned with the laser polarization.  The separate distributions are
much more structured and show their maxima at different
angles. Particularly noteworthy are the pattern from the $1p_x$ and
$1p_y$ states.  Both have their emission maxima at polar angle
$\theta=45^\circ$ and $135^\circ$ and both show pronounced
structures in azimuthal angle $\phi$; the azimuthal pattern is
shifted by $90^\circ$ between $1p_x$ and $1p_y$. That feature will
play a role again for deposited Na$_8$, see section \ref{sec:depos}.
The maxima at $\theta=45^\circ$ and $135^\circ$ can be qualitatively
understood in a picture using wave functions of a spherical mean
field. The $1p_{xy}$ have then the angular wave function in terms of
the spherical harmonics $Y_1^{\pm 1}$. The dipole excitation comes with
angular distribution $Y_1^0$. For $1p_x$, the emitted wave
is driven by the product $|(Y_1^{+1}+Y_1^{-1})Y_1^0|^2\propto
\sin^2(2\theta) \sin^2 \phi$, having maxima at $\phi=0^\circ$ and
$90^\circ$, and for $1p_y$ by $|(Y_1^{+1}-Y_1^{-1})Y_1^0|^2\propto
\sin^2(2\theta) \cos^2 \phi$ with maxima
at $\phi=45^\circ$ and $135^\circ$, while both waves are maximized at
$\theta=45^\circ$ and $135^\circ$. This picture agrees with the
obtained patterns in figure \ref{fig:ang8orbital_019}.
Note that the total cross-section of free Na$_8$ does not exhibit any
structure in $\phi$ as the $1p_x$ and $1p_y$ distributions add up to a
constant value in $\phi$ (besides a faint perturbation by the ions). A
similar finding holds for the energy-resolved distributions as the
states $1p_x$ and $1p_y$ have exactly the same single-electron energy
and thus are added up with equal weight.  We also separately explored
the frequency dependence of the distributions for emission from each
single-electron state. They turned out to depend much less on the
frequency than the total PAD. We refrain from presenting these results
in detail as this would require quite a bit space.
\begin{figure}[htbp]
\centerline{
\includegraphics[width=70mm]{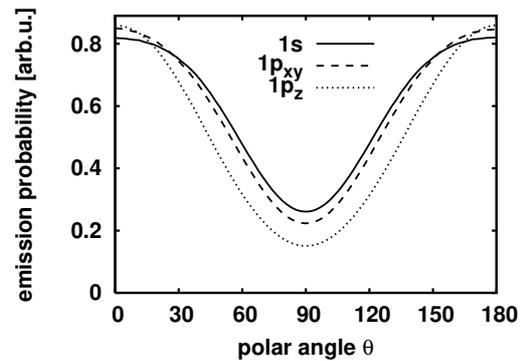}
}
\caption{\label{fig:angdist-av-polar-sp}
Angular distributions, averaged over orientations,
along polar angle $\theta$
for emission from the different single-electron states of
free Na$_8$. 
The laser frequency is  $\omega_\mathrm{las}=5.44$~eV.
}
\end{figure}
Considering free clusters with known orientation allows one to
investigate characteristic structural features and to provide basic
information for deposited clusters to be discussed later on.  An
alignment of clusters in the gas phase remains an open experimental
problem (``orientation burning'' may be an option \cite{Rei05d}).
Present-days experiments in the gas phase measure an ensemble of
clusters where all orientations are equally distributed. A simulation
of that situation requires one to calculate an ensemble of
orientations and to average the emerging angular distributions.  The
procedure is straightforward. We computed angular
distributions on a grid of orientations with grid spacing of
22.5$^\circ$ in Euler angles $\theta$ and $\phi$, ignoring $\psi$ due
to the near axial symmetry of Na$_8$, and added them with appropriate
geometrical weights.
The orientation averaging wipes out the sub-structures along
$\phi$ direction. 
Figure \ref{fig:angdist-av-polar-sp} shows the
resulting (orientation averaged)
distributions along polar angle $\theta$. Note that we have
also averaged over $1p_x$ and $1p_y$ distributions since these two states
are energetically degenerate and could not be discriminated by
PES. 
The orientation averaging yields much different and simpler
structures than those seen in the case of known orientation, see figure
\ref{fig:ang8orbital_019}. The dominance of purely forward-backward
structure holds only for the small system Na$_8$. Larger
clusters show richer structures \cite{Kos07a,Bae08b,Bar08a}.

\section{Emission patterns for deposited Na$_8$}
\label{sec:depos}

\subsection{Deposition on MgO(001)}
\label{sec:deposMg}
\begin{figure}
\centerline{
{\includegraphics[width=80mm]{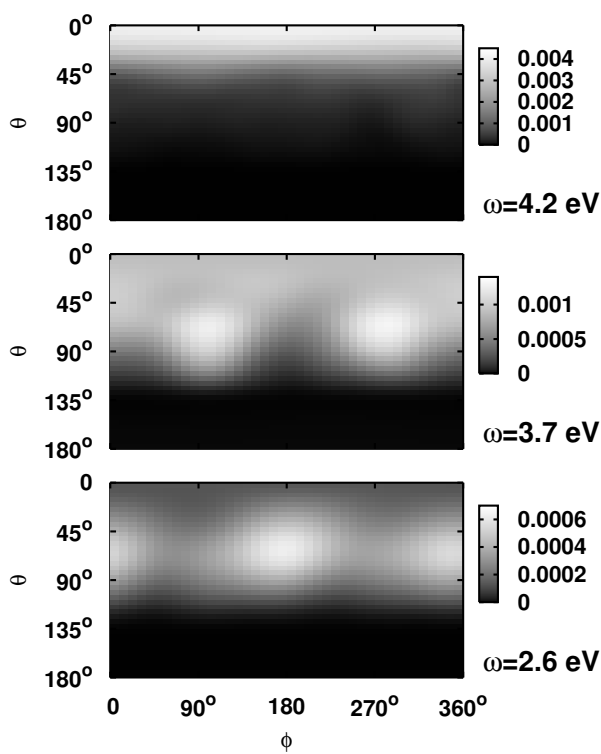}}
}
\caption{\label{fig:typNa8M}
Angular distributions for emission from Na$_8$ on MgO(001) for three 
different laser frequencies as indicated.
The laser is polarized along the $z$-axis, perpendicular to
the surface.
}
\end{figure}
As pointed out in section \ref{sec:free}, the detailed analysis of free
PAD leads to remarkable insights into the structures of
clusters. However, the orientation problem inherent to clusters in the
gas phase blurs some of the detailed information. It is thus
interesting to consider the complementing case of deposited clusters
where the orientation is well defined. On the other hand, the presence
of the substrate complicates the PAD. Analysis of this effect is thus a key
issue.

Figure \ref{fig:typNa8M} displays angular distributions of Na$_8$ on
MgO(001) for three different frequencies. 
{Again, the frequencies were selected to show
three different and typical emission patterns. The sequence
is comparable to the case of free Na$_8$ discussed above, but
having somewhat lower values to accommodate for the lower IP
(see figure  \ref{fig:Na8-spe}) and stay below the
transmission threshold.}
The substrate obviously has
a very strong influence such that the patterns are quite different from
those of free Na$_8$.
The dominant forward emission ($\theta = 0^\circ$) is still observed;
backward emission ($\theta = 180^\circ$) is, of course, totally
suppressed by the presence of the insulating substrate. More
surprising is the appearance of a strong azimuthal dependence.  
Recall the pronounced azimuthal structures for emission from the
$1p_{x,y}$ levels of free (aligned) clusters, see figure
\ref{fig:ang8orbital_019}.  In the total or energy-resolved
cross-sections of free clusters, these structures are wiped out by the 
degeneracy, as each of the two levels contributes equally to the
emission strength, see the discussion of figure
\ref{fig:ang8orbital_019}. This $1p_x$--$1p_y$ degeneracy is now
split by the surface leading to an energy difference between $1p_x$
and $1p_y$ state of 0.2 eV. Because these states are close to the
emission threshold, the small energy difference has a large effect on
the relative emission strengths.  One of the two states dominates
emission and its pattern affects the total distribution.  Comparison
to figure \ref{fig:ang8orbital_019} makes it clear that the dominant
state is $1p_x$ for $\omega_\mathrm{las}=2.6$~eV (figure
\ref{fig:typNa8M},  lower panel) and
$1p_y$ for $\omega_\mathrm{las}=3.7$~eV (figure \ref{fig:typNa8M},
middle panel).
There is also some effect on the polar distribution.  The emission
cones are neither perfectly aligned along $\theta=90^\circ$, as in a
sidewards emitting case,
nor along $\theta=45^\circ$, 
as was the case for $1p_{x,y}$ orbitals of free Na$_8$.
The composition of emission strengths from the four single electron
levels is different for the deposited case as compared to free
Na$_8$. Furthermore, the polarization attraction bends the outwards
cones towards the substrate. Both effects together widen the emission
cone such that the angle of inclination of the emitted electrons
relative to the surface decreases.

This example demonstrates that the detailed structure of the angular
distribution, in the polar as well as in the azimuthal angles,
sensitively depends on the laser frequency. There is much more
structure compared to the emission spectra of free clusters, thus more
information, worth to be unraveled. That information, however, is
masked by the involved interplay of cluster and surface which inhibits
simple one-to-one correspondences and requires a careful correlation
analysis.

\subsection{Comparison with argon substrate}

We also investigated the case for another combination, namely Na$_8$
on Ar(001). Frozen Ar is an insulator as MgO is. We thus expected, as
in the case of Na$_8$ on MgO(001), heavily modified angular
distributions compared with the free Na$_8$ case. Ar is, however, a
Van-der-Waals bound system with a smaller surface corrugation and
smaller initial polarization fields, in contrast to the ionic crystal
MgO.
\begin{figure}[htbp]
\centerline{
{\includegraphics[width=80mm]{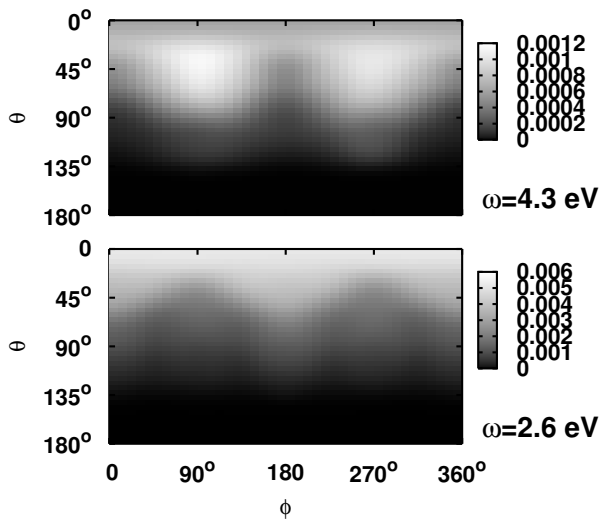}}
}
\caption{\label{fig:PES_Argon}
Angular distributions for emission
from Na$_8$ on Ar(001) for two laser frequencies as indicated.
The laser is polarized along the $z$-axis, perpendicular to
the surface.
}
\end{figure}
Figure \ref{fig:PES_Argon} shows angular distributions for Na$_8$ on
Ar(001) at two frequencies of the laser field.  The effects are very
similar to the case of MgO. Backward scattering is totally suppressed,
forward scattering is accordingly enhanced, there appear pronounced
azimuthal structures, and the distributions depend sensitively on the
frequency. With $0.14\,\rm eV$, the splitting of the 1$p_{x,y}$ levels
due to Ar(001) surface is similar to that at MgO, which is probably
responsible for the similarities of the emission patterns between Ar
and MgO.
There is, however, one difference to the case of MgO. The sideward
cone (lower panel of figure \ref{fig:PES_Argon}) is not so close to
the surface and points towards a polar angle of $\theta=45^\circ$ as
in the case of free Na$_8$. That is probably due to the smaller
surface polarization for Ar.

\subsection{Frequency and intensity dependences}

In this section we explore more systematically the
influence of laser frequency and intensity on the angular
distributions for deposited Na$_8$.  
To simplify the graphical representation, we consider 
$\phi$-averaged distribution along polar angle $\theta$.
\begin{figure}[htbp]
\centerline{
\includegraphics[width=70mm]{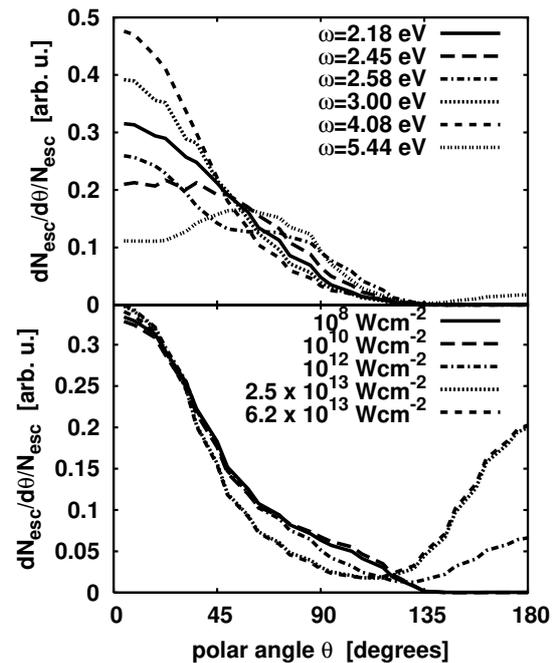}
}
\caption{\label{fig:angtheta8M}
Top~: Azimuthally averaged angular distribution 
(arbitrary units) of electrons emitted from
Na$_8$ on MgO(001) irradiated by laser pulses with intensity
$I=10^9$ W/cm$^2$ and varying frequency.
The laser is polarized along the $z$-axis, perpendicular to
the surface.
Bottom~: Distribution for fixed frequency $\omega_{\rm las}=4.76\,{\rm
  eV}$ with varying laser intensity.
}
\end{figure}
Figure \ref{fig:angtheta8M} shows azimuthally averaged
distributions for emission from Na$_8$ on MgO(001).
The upper panel presents the variation with $\omega_\mathrm{las}$ for
moderate intensity. The prevailing effect is the strong suppression of
backward emission through the substrate. In spite of reflection-enhanced
surface dominance, we see again the variation of pattern with
frequency qualitatively similar to the case of free Na$_8$
(see figure \ref{fig:angdist-frequency}).  The pattern is,
however, more involved than what would emerge from a simple rescattering of
the reflected electrons to angle
$\theta\longrightarrow 180^\circ-\theta$. The surface attraction shifts the
former diagonal maximum from $45^\circ$ towards the substrate and
produces sizeable emission under rather flat angles
$\theta\approx 90^\circ$. A first glimpse of transmission into the
substrate can be seen for $\omega_\mathrm{las}=5.44$ eV.
That is very close to the transmission threshold for Na$_8$ on Mg(001)
at about 5.58~eV, see figure \ref{fig:Na8-spe}.

The lower panel of figure \ref{fig:angtheta8M} shows the changes with
increasing intensity for fixed frequency. We see again the expected
typical increasing focus towards forward emission. A somewhat
surprising and most interesting effect is that backward emission
gathers rather large strength with further increasing
intensity. The high field strengths make two- and more-photon
processes competitive which, in turn overrules the frequency counting.
However, one has to be cautious with interpreting these processes where
substantial amounts of electron charge are penetrating into the
substrate.  The present QM/MM model is not set up for that situation
and likely is no longer accurate at a quantitative level.


\begin{figure}[htbp]
\centerline{
{\includegraphics[width=70mm]{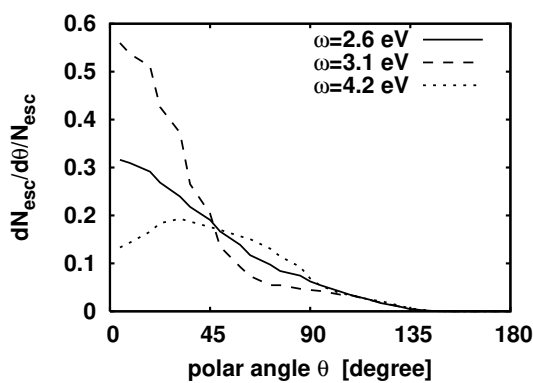}}
}
\caption{\label{fig:PESpolar_Argon}
Azimuthally averaged angular distributions 
(arbitrary units) for Na$_8$ on Ar(001) for
intensity $I=10^9$ W/cm$^2$ and three different laser frequencies, as
indicated.  The laser is polarized along the $z$-axis, perpendicular to
the surface.
}
\end{figure}
Figure \ref{fig:PESpolar_Argon} shows some azimuthally averaged
distributions for Na$_8$ on Ar(001).  The general trends are very
similar to the case of MgO, see for example  the upper panel of figure
\ref{fig:angtheta8M}. We see again the strong forward dominance due
to reflection from the substrate and the strong dependence of the
distribution on laser frequency as was observed before for all cases.
%
There is a minor difference in that the chances for transmission into
the substrate are even slightly lower.

\section{Conclusion}
\label{sec:concl}

We have explored from a theoretical perspective angular distributions
of electrons emitted from Na$_8$ clusters, free and deposited on
Ar(001) or MgO(001) surfaces. Thereby we concentrated on direct
electron emission which takes place within few fs
after laser excitation and which is the dominant process for short
laser pulses (up to about 50 fs).  Several laser frequencies were
used, higher ones above the threshold for one-photon emission, lower
frequencies in the two-photon regime close to the Mie
plasmon resonance, and cases safely off-resonance.  The laser intensity
was in most cases $I=10^9$ W/cm$^2$, large enough to overcome
numerical noise, but safely below the regime of violent and highly
non-linear processes.
The numerical simulations employed a hierarchical
quantum-mechanical-molecular-mechanical (QM/MM) model approach.
The cluster electrons were described by time-de\-pen\-dent
density-functional theory (TDDFT) at the level of the local-density
approximation (LDA). A self-interaction correction has been added to put the
single-electron levels to the appropriate energy relative to the
particle continuum. The cluster ions were propagated by classical
molecular dynamics. Similarly, the Ar(001) or MgO(001) substrates
were treated as classical systems with atomic positions and dipole
polarizability as dynamical degrees of freedom. The TDDFT equations
were solved on a three-dimensional co\-ordinate-space grid without any
symmetry restriction. Absorbing boundaries were applied and the
angular distributions were obtained by recording the electron
absorption at each grid point in the absorbing zone.

We first briefly explored free Na$_8$.  The state-specific
distributions for Na$_8$ aligned with laser polarization show
pronounced patterns in both angles, $\theta$ and $\phi$, which exhibit
clear footprints of the emitting state.
Simulating the experimental situation which deals with orientation
averaged ensembles renders the distributions independent of the
azimuthal angle $\phi$ and reduces structures in the polar angles
$\theta$.  A study of averaging effects, size-, frequency-, and
intensity-dependence for free clusters will follow in an upcoming
publication.

The present study focused on angular distributions from Na$_8$
deposited on MgO(001) and Ar(001). The attachment to the surface
provides a well defined cluster orientation and it adds substantial
perturbations from the surface interaction.  The large band gaps of
both materials manifested in high transmission barriers, almost
totally suppress backward emission, focusing electrons into a forward
cone. On the other hand, the long-range polarization attraction can
bend sideward flowing electrons down towards the surface, thus
widening the opening angle of the emission cone. Polarization is
strong for MgO(001) which induces a sizeable trend to emission
parallel to the surface. Repulsion dominates for Ar(001) which
stabilizes the forward cone. The surfaces do also break the fourfold
symmetry ($C_4$) of Na$_8$ which, in turn, removes the degeneracy of
the two single-electron levels with nodes orthogonal to the symmetry
axis, and thus inhibits the compensation of $\phi$ dependence which
was observed for (oriented) free Na$_8$. Pronounced azimuthal
structures are found for the distributions of the deposited cases.

Altogether, the computational study revealed interesting structures
which are, however, somewhat complicated by surface (particularly the
electronic and geometric structure of the clusters) contained in the
calculated photoelectron angular distributions and their energy
dependence.

\begin{acknowledgement}
This work was supported by the 
{Deutsche Forschungsgemeinschaft (RE 322/10, RO 293/27),}
Fonds der Chemischen Industrie (Germany), a Bessel-Humboldt prize, and
a Gay-Lussac prize.
\end{acknowledgement}

\bibliographystyle{pss}
\bibliography{laser2}

\end{document}